\documentclass{ecai} 

\usepackage{latexsym}
\usepackage{times}
\usepackage{soul}
\usepackage{url}
\usepackage[hidelinks]{hyperref}
\usepackage[utf8]{inputenc}
\usepackage[small]{caption}
\usepackage{graphicx}
\usepackage{amsmath}
\usepackage{amsthm}
\usepackage{booktabs}
\usepackage{algorithm}
\usepackage{algorithmic}
\usepackage[switch]{lineno}
\usepackage{natbib}
\usepackage{amsfonts}   
\usepackage{amssymb}
\usepackage{array}
\usepackage{xcolor}
\usepackage{tabularx}
\usepackage{adjustbox}
\usepackage{multirow}
\usepackage[utf8]{inputenc}
\usepackage{threeparttable}
\usepackage[table]{xcolor}

\usepackage{latexsym}
\usepackage{amssymb}
\usepackage{amsmath}
\usepackage{amsthm}
\usepackage{booktabs}
\usepackage{enumitem}
\usepackage{graphicx}
\usepackage{color}

\usepackage{graphicx}
\usepackage{latexsym}

\usepackage{times}
\usepackage{soul}
\usepackage{url}
\usepackage[hidelinks]{hyperref}
\usepackage{amsthm}
\usepackage{booktabs}
\usepackage{algorithm}
\usepackage{algorithmic}
\usepackage[switch]{lineno}

\usepackage{textcomp}
\usepackage{xcolor}
\usepackage{subfigure}
\usepackage{diagbox}
\usepackage{threeparttable}
\usepackage{multirow}
\usepackage{color}
\usepackage{bbold}
\usepackage{setspace}
\usepackage{fontawesome}
\usepackage{fancyvrb}

\newcommand{\BibTeX}{B\kern-.05em{\sc i\kern-.025em b}\kern-.08em\TeX}

\begin{document}


\begin{frontmatter}


\paperid{123} 


\title{AI Fairness Beyond Complete Demographics: Current Achievements and Future Directions}

\author[A]{\fnms{Zichong}~\snm{Wang}}
\author[A]{\fnms{Zhipeng}~\snm{Yin}}
\author[B]{\fnms{Roland}~H. C. \snm{Yap}}
\author[A]{\fnms{Wenbin}~\snm{Zhang}\thanks{Corresponding author. Email: wenbin.zhang@fiu.edu}}

\address[A]{Florida International University, Miami, Florida, 33199}
\address[B]{National University of Singapore, Singapore, 117417}





\begin{abstract}
 Fairness in artificial intelligence (AI) has become a growing concern due to discriminatory outcomes in AI-based decision-making systems. While various methods have been proposed to mitigate bias, most rely on complete demographic information, an assumption often impractical due to legal constraints and the risk of reinforcing discrimination. This survey examines fairness in AI when demographics are incomplete, addressing the gap between traditional approaches and real-world challenges. We introduce a novel taxonomy of fairness notions in this setting, clarifying their relationships and distinctions. Additionally, we summarize existing techniques that promote fairness beyond complete demographics and highlight open research questions to encourage further progress in the field.
\end{abstract}

\end{frontmatter}


\section{Introduction}

Artificial intelligence (AI) based decision-making systems are widely deployed in real-world applications, but concerns about their potential to produce discriminatory outcomes have raised significant ethical and social challenges, particularly in high-stakes domains such as healthcare, finance, and criminal justice~\citep{pelegrina2023statistical,pmlr-v267-wang25ek,ijcai2025p63,wang2026guic}. In response, research on fairness in AI has gained increasing attention, with a primary focus on statistical group fairness approaches~\citep{hardt2016equality}. These methods operate under the complete demographic assumption, defining a limited set of groups, such as those based on gender or race, and enforcing similar outcome statistics, including prediction accuracy and true positive rates, across these predefined groups to mitigate systemic disparities in decisions such as medical care allocation and targeted marketing.

In practice, however, this assumption often fails in real-world situations due to privacy, legal, and regulatory constraints. For instance, the EU GDPR restricts the collection of personal data revealing demographics such as racial or ethnic origin, political opinions, religious beliefs, or trade union membership~\citep{voigt2017eu}. In addition, individuals may choose not to share their gender in health surveys due to privacy concerns or perceived irrelevance~\citep{ogden2020feel}, while others might withhold this information when applying for jobs in fields dominated by the opposite gender, fearing discrimination~\citep{michie2006barriers}. These restrictions undermine the effectiveness of existing fairness methods that assume complete demographics, leaving users facing socially sensitive issues without adequate tools to address concerns of discrimination~\citep{chai2022fairness}.

To address this gap, recent efforts have explored approaches for handling incomplete demographic information, yet fairness beyond the assumption of complete demographics remains an emerging research area. Despite numerous surveys on AI fairness under complete demographic assumptions~\citep{le2022survey}, there is a striking lack of studies investigating how fairness can be ensured when demographics are incomplete. To this end, \textit{this paper presents, to the best of our knowledge, the first comprehensive survey examining which fairness concepts are effective in these settings and analyzing the challenges, assumptions, and limitations of existing methods}. By extending prior fairness surveys with a critical focus on incomplete demographic information, this work advances the field in several key ways. Specifically, this survey makes four contributions: \textbf{i)} it introduces a novel taxonomy of fairness under incomplete demographic information, providing formal definitions and quantitative metrics for each fairness type; \textbf{ii)} it offers a systematic review of six major methodological approaches for achieving fairness in such settings; \textbf{iii)} it critically evaluates research gaps, identifies key challenges, and outlines open questions to guide future advancements in this domain; and \textbf{iv)} it collects rich resources of benchmark datasets that can be employed for fairness research involving missing demographic information, thus facilitating the development of new approaches to promote fairness.

The remainder of this survey is organized as follows. 
Section~\ref{sec:Notation} presents notation and background. Section~\ref{sec:notions} systematically reviews different fairness notions for incomplete demographic information. Section~\ref{sec:methods} introduces six groups of techniques that implement these fairness concepts. Section~\ref{sec:datasets} describes commonly used datasets. Section~\ref{sec:challenge} discusses research challenges and open questions. Finally, Section~\ref{sec:conclusion} concludes the survey.

\section{Notation and Background}
\label{sec:Notation}

\noindent \textbf{Notation.} We denote matrices with bold uppercase letters (\textit{e.g.}, $\mathbf{A}$), and vectors by bold lowercase letters (\textit{e.g.}, $\mathbf{z}$). For a matrix $\mathbf{A}$, the element in the $i^{th}$ row and $j^{th}$ column is indicated by $\mathbf{A}_{i,j}$. \textbf{i)} For I.I.D. data, samples are represented as $(x_i, y_i, s_i)$ drawn from the domain $\mathbf{X} \times \mathbf{Y} \times \mathbf{S}$, where each feature vector $x_i \in \mathbb{R}^{d}$, $y_i$ denotes the ground truth label, and $s_i$ is demographic information. \textbf{ii)} For non-I.I.D. data, samples are structured as a graph $\mathcal{G} = (\mathcal{V}, \mathcal{E})$, represented by an adjacency matrix $\mathbf{A} \in {0,1}^{n \times n}$, where $n$ is the number of nodes, and $\mathbf{A}_{i,j}=1$ indicates an edge between nodes $v_i$ and $v_j$. Nodes correspond to individuals or entities, each described by features, demographics $s_i$, and labels $y_i$.

\noindent \textbf{Demographic Information.} Demographic information consists of attributes such as race, gender, and age, which may classify individuals into groups that are sometimes treated differently. These features are often protected by laws and international guidelines to prevent discrimination. Many regulations prohibit using these protected attributes as a basis for decision-making to avoid unfair treatment. For example, the Consumer Financial Protection Bureau (CFPB) requires creditors to ensure fairness without collecting or using information about an applicant’s race, color, religion, national origin, or sex~\citep{chai2022fairness}. In such cases, demographic information is completely unavailable. In other situations, individuals choose not to share their demographic information. For example, only 14\% of teen users share their complete profiles on Meta~\citep{madden2013teens}, making demographic information partially available.

\section{Fairness Notions Beyond Complete Demographics}
\label{sec:notions}

This section introduces a new taxonomy that categorizes fairness notions with incomplete demographics into six types: traditional Rawlsian fairness, group fairness, counterfactual fairness, proxy fairness, individual fairness, and fairness under unawareness. Our categorization follows established fairness principles~\cite{zhang2025fairness} but is organized along two key dimensions: i) whether they provide group-level protection or individual-level protection; and ii) whether they rely on explicit demographic information, latent demographic proxies, or entirely demographic-free reasoning. This organizational structure helps practitioners select appropriate fairness approaches based on their specific constraints regarding demographic data availability and their fairness goals. The following subsections describe these notions along with their corresponding metrics.

\subsection{Rawlsian Fairness}

Rawlsian Fairness is based on John Rawls’s difference principle~\citep{rawls1971theories}, which aims to improve the well-being of the least advantaged group. Unlike group fairness notions that rely on demographic information, Rawlsian Fairness focuses on identifying the least advantaged group through the decision-making process itself~\cite {rawls2001justice}. 
The system reaches equilibrium when the welfare of this group cannot improve further without reducing the welfare of other groups. Formally, let $\mathbf{X}$ denote the feature space and $\mathbf{Y}$ denote the decision outcome space (\textit{e.g.}, loan approval or rejection). A decision function $f_{\theta}:\mathbf{X}\rightarrow\mathbf{Y}$, parameterized by a model parameter vector $\theta \in \Theta$, maps individual features $x \in \mathbf{X}$ to a decision outcome $y \in \mathbf{Y}$. We let $\Theta$ represents the set of all possible parameters of the decision-making model. Let $\{G_1,\dots,G_k\}$ denote $k$ sub-populations identified by the decision maker. For each subgroup $G_s$, we define the mean utility:

\begin{equation}
    U(G_{s};f_{\theta}) \;=\;
    \mathbb{E}_{(x,y)\in G_{s}}
    \!\bigl[\,u\bigl(f_{\theta}(x),y\bigr)\bigr]   
\end{equation}

\noindent where $u(\cdot,\cdot)$ is a utility function that measures the benefit or satisfaction derived from decisions.

Rawlsian fairness is then obtained by minimizing the variance of these subgroup utilities:

\begin{equation}
\label{eq:rawlsian}
\min_{\theta\in\Theta}\;
\operatorname{Var}_{s=1,\dots,k}
\!\Bigl[\,U\bigl(G_{s};f_{\theta}\bigr)\Bigr]
\end{equation}

When this variance reaches its minimum, all subgroups achieve equal average utility. Consequently, any further improvement in the welfare of the least-advantaged subgroup would equally benefit all other subgroups, aligning with Rawls' maximin intuition.

Although this notion works without directly using demographic information, it has limitations in implementing Rawlsian distributive justice. In classification tasks, the least advantaged group might primarily consist of misclassified instances. Improving performance only for these cases may fail to address truly disadvantaged communities and can make the model highly sensitive to outliers~\citep{hashimoto2018fairness,lahoti2020fairness}. Additionally, because Max-Min fairness typically requires similar utility across all groups, using an excessive number of subpopulation groups can significantly degrade model performance~\citep{chai2022fairness}.

On the other hand, unlike group fairness notions, which reduce disparities in specific metrics, Rawlsian fairness seeks to maximize the outcomes of the worst-performing group. It can also handle intersectional groupings and be applied to tasks beyond classification, provided that a suitable measure of utility is available.

\subsection{Group Fairness}

Group fairness~\citep{hardt2016equality} ensures that algorithmic decisions do not unfairly disadvantage certain groups defined by demographic information, such as gender, race, or age. This approach begins by explicitly identifying or predefining demographic characteristics that could lead to biased outcomes, and then measures fairness based on disparities in utility or performance outcomes (such as differences in accuracy, acceptance rates, or true positive rates) across privileged and disadvantaged groups. Specifically, fairness metrics typically quantify these disparities through measures such as demographic parity~\citep{dwork2012fairness}, which requires equal acceptance or positive decision rates across demographic subgroups, or equality of opportunity~\citep{hardt2016equality}, which demands that qualified individuals from different groups have equal probabilities of favorable outcomes (\textit{e.g.}, being granted a loan).

However, applying these fairness measures directly requires complete demographic information at training and evaluation stages. In practice, fully demographic information is often difficult to obtain due to privacy regulations, data protection laws, or individuals' unwillingness to disclose demographic information. This limitation affects numerous fairness criteria beyond demographic parity and equality of opportunity, including Equalized Odds~\cite{hardt2016equality} and Equal Opportunity Difference~\cite{chakraborty2021bias}, all of which rely on explicitly available demographic information. Despite these challenges, group fairness remains essential, motivating ongoing research into alternative methods that can maintain fairness even when demographic data is incomplete or unavailable.

In scenarios lacking explicit demographic information, proxy demographics can often be inferred or approximated using related non-sensitive attributes or observable patterns. Such proxy variables might include geographic indicators, behavioral characteristics, or other indirect signals correlated with demographic groups. Utilizing these inferred proxy demographics enables researchers and practitioners to estimate fairness metrics at evaluation time, providing insights into potential biases even in the absence of complete demographic information~\citep{liang2023fair,wang2025towards}. Although these inferred demographics may introduce some approximation error or uncertainty, they remain valuable for assessing fairness and guiding decisions around algorithmic interventions.

\subsection{Counterfactual Fairness}

Counterfactual fairness~\citep{kusner2017counterfactual} defines fairness from a causal perspective, which differs from statistical fairness definitions (\textit{e.g.}, group fairness). Specifically, counterfactual fairness requires that a classifier's prediction for an individual remain consistent even if the individual's demographic information is altered in a counterfactual scenario. For instance, a fair loan application process should yield the same decision for a particular individual, whether the applicant's race or gender were to differ in a counterfactual scenario, provided all other attributes remain constant. Formally, for any demographic information $S$, decision outcome $\hat{Y}$, and observed features $X$, counterfactual fairness is satisfied if:

\begin{equation}
P(\hat{Y}_i(S \leftarrow s) \mid X, S) = P(\hat{Y}_i(S \leftarrow s') \mid X, S)
\end{equation}

\noindent where $s$ and $s'$ represent any two different values from the set of all possible demographic values $\mathcal{S}$. This definition emphasizes a causal interpretation of fairness by explicitly incorporating causal relationships among features and decisions, typically represented through structural causal models (SCMs)~\cite{pearl2009causality}.

Recent research~\cite{zhang2021multi} has introduced counterfactual fairness variants, particularly those using the Unfairness Score, which quantifies the fraction of individuals whose predicted labels would change if their demographic information were altered counterfactually, keeping all other factors fixed~\cite{zhang2025fairness}. Such measures enable empirical evaluation of a model's susceptibility to discriminatory biases that depend causally on demographic information. However, similar to group fairness notions, these variants cannot be directly applied when demographic information is incomplete. The following presents fairness definitions designed to function without complete demographics.

\renewcommand{\arraystretch}{1.5}

\begin{table*}
    \centering
    \caption{Summary of various methodologies used for fairness beyond complete demographics.}
    \vspace{+0.2cm}
    \begin{adjustbox}{width=0.88\textwidth}
    \begin{tabular}{c| c| c| c| c| c| c}
        \toprule
        \toprule
        \multirow{2}{*}{\textbf{Fairness Type}}  & \multirow{2}{*}{\textbf{Method}} &\multirow{2}{*}{\textbf{Missingness}}& \multirow{2}{*}{\textbf{Data Type}} & \multirow{2}{*}{\textbf{Class}} & \multirow{2}{*}{\textbf{Evaluation Metrics}} & \multirow{2}{*}{\textbf{Remark}} \\
        &&&&&&\\
        \midrule
        \midrule
        \multirow{16}{*}{Rawlsian Fairness}  & WGA&\multirow{2}{*}{Completely}  &\multirow{2}{*}{Non-I.I.D.} &\multirow{2}{*}{In-processing} & Average Accuracy, & \multirow{2}{*}{Confidence weighting}\\
          & ~\citep{tiwari2024using} & & &    & WGA  &\\
          
        \cmidrule{2-7}
        
          & \cellcolor{gray!25}ECS&\cellcolor{gray!25} &\cellcolor{gray!25}  &\cellcolor{gray!25} & \cellcolor{gray!25}Accuracy, DP, & \cellcolor{gray!25}\\
         &\cellcolor{gray!25}~\citep{zhao2023combating} &\cellcolor{gray!25}\multirow{-2}{*}{Completely}  &\cellcolor{gray!25}\multirow{-2}{*}{I.I.D.} & \cellcolor{gray!25} \multirow{-2}{*}{In-processing}   & \cellcolor{gray!25}EOD & \cellcolor{gray!25}\multirow{-2}{*}{Gradient alignment}\\
         
        \cmidrule{2-7}
           & DRO-COX& \multirow{2}{*}{Completely}&\multirow{2}{*}{I.I.D.} &\multirow{2}{*}{In-processing} & C-index, AUROC,  & \multirow{2}{*}{Survival analysis} \\
          & ~\citep{hu2022distributionally}& & &    &  LPL, IBS, $F_A$,  CI & \\
          
        \cmidrule{2-7}
        
          & \cellcolor{gray!25}BPF&\cellcolor{gray!25} &\cellcolor{gray!25} &\cellcolor{gray!25} & \cellcolor{gray!25}Worst Group Cross-entropy, & \cellcolor{gray!25}\\
         & \cellcolor{gray!25}~\citep{martinez2021blind}& \cellcolor{gray!25}\multirow{-2}{*}{Completely}& \multirow{-2}{*}{\cellcolor{gray!25}I.I.D.}& \cellcolor{gray!25}\multirow{-2}{*}{In-processing} &  \cellcolor{gray!25}Worst Group Error rate  & \multirow{-2}{*}{\cellcolor{gray!25}Pareto optimality}\\
        \cmidrule{2-7}
        
           & groupDRO& \multirow{2}{*}{Completely}&\multirow{2}{*}{I.I.D.} &\multirow{2}{*}{In-processing} & Average Accuracy, &  \multirow{2}{*}{$\chi^2$-divergence}\\
         & ~\citep{sagawa2019distributionally} &  & &  & Worst-Group Accuracy & \\
         
        \cmidrule{2-7}
        
           & \cellcolor{gray!25}FairKD&\cellcolor{gray!25} &\cellcolor{gray!25} &\cellcolor{gray!25} & \cellcolor{gray!25}Accuracy, DI, & \cellcolor{gray!25} \\
          & \cellcolor{gray!25}~\citep{chai2022fairness}& \cellcolor{gray!25}\multirow{-2}{*}{Completely}&  \cellcolor{gray!25}\multirow{-2}{*}{I.I.D.}& \cellcolor{gray!25}\multirow{-2}{*}{In-processing}&  \cellcolor{gray!25} EOD & \cellcolor{gray!25}\multirow{-2}{*}{Multiple sensitive attributes}\\
          
        \cmidrule{2-7}
        
          & DRO&\multirow{2}{*}{Completely} &\multirow{2}{*}{I.I.D.} &\multirow{2}{*}{In-processing} & Accuracy, Minority accuracy,  &  \multirow{2}{*}{Distributionally robust optimization}\\
          & ~\citep{hashimoto2018fairness}& & &   &User satisfaction and  Retention rate & \\
        \bottomrule
        \toprule
        \multirow{31}{*}{Proxy Demographic}   & \cellcolor{gray!25}Themis&\cellcolor{gray!25}  &\cellcolor{gray!25}  &\cellcolor{gray!25} & \cellcolor{gray!25}Accuracy, F1-score,  & \cellcolor{gray!25}\\
          \multirow{31}{*}{Information}& \cellcolor{gray!25}~\citep{wang2025towards}& \cellcolor{gray!25}\multirow{-2}{*}{Completely}& \cellcolor{gray!25}\multirow{-2}{*}{Non-I.I.D.}& \cellcolor{gray!25}\multirow{-2}{*}{In-processing}& \cellcolor{gray!25}EOD, SPD & \cellcolor{gray!25}\multirow{-2}{*}{Exclude noise information}\\
          
         \cmidrule{2-7}
         
            & Fairwos &\multirow{2}{*}{Completely}&\multirow{2}{*}{Non-I.I.D.} &\multirow{2}{*}{In-processing} & Accuracy, EOD & \multirow{2}{*}{Counterfactual fairness}\\
          & ~\citep{wang2024towards}& &  & &   SPD & \\

         \cmidrule{2-7}
        
           & \cellcolor{gray!25}fairGNN-WOD& \cellcolor{gray!25}&\cellcolor{gray!25} &\cellcolor{gray!25} & \cellcolor{gray!25}Accuracy, F1-score, & \cellcolor{gray!25}\\
          & \cellcolor{gray!25}~\citep{ijcai2025p63}& \cellcolor{gray!25}\multirow{-2}{*}{Completely}&\cellcolor{gray!25}\multirow{-2}{*}{Non-I.I.D.} & \cellcolor{gray!25} \multirow{-2}{*}{In-processing}&  \cellcolor{gray!25}SPD, EOD  & \cellcolor{gray!25}\multirow{-2}{*}{Two-stage framework}\\

          \cmidrule{2-7}

            & KSMOTE&\multirow{2}{*}{Completely} &\multirow{2}{*}{I.I.D.} &\multirow{2}{*}{Pre-processing} & Accuracy, F1-score, & \multirow{2}{*}{Multiple sensitive attributes}\\
          & ~\citep{yan2020fair}& &&  &  EOD, EOP, SPD&   \\

          \cmidrule{2-7}
          
            & \cellcolor{gray!25}NOCOO& \cellcolor{gray!25}&\cellcolor{gray!25} &\cellcolor{gray!25} & \cellcolor{gray!25} & \cellcolor{gray!25}\\
          &\cellcolor{gray!25} ~\citep{pelegrina2023statistical}&  \cellcolor{gray!25}\multirow{-2}{*}{Completely} &\cellcolor{gray!25}\multirow{-2}{*}{I.I.D.}& \cellcolor{gray!25}\multirow{-2}{*}{Pre-processing} &\cellcolor{gray!25} \multirow{-2}{*}{Accuracy, EOD}& \cellcolor{gray!25}\multirow{-2}{*}{ Multiple sensitive attributes}\\
          
        \cmidrule{2-7}
        
           & FairRF& & & & Accuracy, EOD, &  \\
            &  ~\citep{zhao2022towards}& \multirow{-2}{*}{Completely}&  \multirow{-2}{*}{I.I.D.}& \multirow{-2}{*}{In-processing} & SPD  &  \multirow{-2}{*}{Adopting related features}\\  
            
          \cmidrule{2-7}
          
           & \cellcolor{gray!25}SLSD& \cellcolor{gray!25}  &\cellcolor{gray!25} &\cellcolor{gray!25} &\cellcolor{gray!25} Accuracy,AUROC &\cellcolor{gray!25} \\
          &\cellcolor{gray!25} ~\citep{islam2024fairness}&\cellcolor{gray!25}\multirow{-2}{*}{Completely} &\cellcolor{gray!25} \multirow{-2}{*}{I.I.D.} &\cellcolor{gray!25} \multirow{-2}{*}{In-processing} &\cellcolor{gray!25} DPD, DPR &\cellcolor{gray!25} \multirow{-2}{*}{Multiple sensitive attributes}\\
          
        \cmidrule{2-7}
        
           & SRCVAE& & & & Accuracy, EOD, & \\
          & ~\citep{grari2021fairness}& \multirow{-2}{*}{Completely}&\multirow{-2}{*}{I.I.D.} &  \multirow{-2}{*}{In-processing}&  EOP  & \multirow{-2}{*}{Bayesian inference}\\

          \cmidrule{2-7}
          
          &\cellcolor{gray!25} FairAC &\cellcolor{gray!25}&\cellcolor{gray!25} &\cellcolor{gray!25} & \cellcolor{gray!25}Accuracy, AUROC, &\cellcolor{gray!25} \\
          &\cellcolor{gray!25}~\citep{guo2023fair}  &\cellcolor{gray!25}  \multirow{-2}{*}{Partially} &\cellcolor{gray!25} \multirow{-2}{*}{Non-I.I.D.}& \cellcolor{gray!25}\multirow{-2}{*}{Pre-processing} & \cellcolor{gray!25}EOD, SPD & \cellcolor{gray!25}\multirow{-2}{*}{Complete attribution completion}\\

          \cmidrule{2-7}
        &DFGR & & & & Accuracy, F1-score,  & \\
          & ~\citep{wang2025Fairness}&\multirow{-2}{*}{Partially} &\multirow{-2}{*}{Non-I.I.D.} & \multirow{-2}{*}{In-processing}& $\Delta_{\rm{DP}}$, $\Delta_{\rm{EO}}$  & \multirow{-2}{*}{Adaptivity Confidence Strategy}\\
          
        \cmidrule{2-7}

         & \cellcolor{gray!25}FGLISA& \cellcolor{gray!25}&\cellcolor{gray!25} &\cellcolor{gray!25} & \cellcolor{gray!25}Accuracy, F1-score, & \cellcolor{gray!25}\\
          & \cellcolor{gray!25}~\citep{wang2026fair}& \cellcolor{gray!25}\multirow{-2}{*}{Partially}&\cellcolor{gray!25}\multirow{-2}{*}{Non-I.I.D.} & \cellcolor{gray!25} \multirow{-2}{*}{In-processing}&  \cellcolor{gray!25}SPD, EOD  & \cellcolor{gray!25}\multirow{-2}{*}{Prevent Manipulation}\\
          
          \cmidrule{2-7}

              & FairDSR&\multirow{2}{*}{Partially} &\multirow{2}{*}{I.I.D.} &\multirow{2}{*}{Pre-processing} & Accuracy, EOD, & \multirow{2}{*}{Uncertainty-aware sensitive attribute classifier} \\
          &  ~\citep{aguirre2023transferring}& &  &  &EOP, SPD  &  \\

        \cmidrule{2-7}
        
             & \cellcolor{gray!25}CGL&\cellcolor{gray!25}&\cellcolor{gray!25} &\cellcolor{gray!25} & \cellcolor{gray!25} &   \cellcolor{gray!25}\\
          &\cellcolor{gray!25} ~\citep{jung2022learning}&\cellcolor{gray!25}\multirow{-2}{*}{Partially}  & \cellcolor{gray!25} \multirow{-2}{*}{I.I.D.}& \cellcolor{gray!25} \multirow{-2}{*}{Pre-processing}& \cellcolor{gray!25}\multirow{-2}{*}{Accuracy, EOP} & \cellcolor{gray!25}\multirow{-2}{*}{Confidence-based attribute classifier} \\
          
        \cmidrule{2-7}
          &FDVAE &\multirow{2}{*}{Partially} & \multirow{2}{*}{I.I.D.}  &  \multirow{2}{*}{In-processing}& Accuracy, F1-score,  & \multirow{2}{*}{Disentanglement framework}\\
        & ~\citep{ijcai2025p64}& & & &$\Delta_{\rm{DP}}$, $\Delta_{\rm{EO}}$  & \\

        \cmidrule{2-7}

        &\cellcolor{gray!25}APOD &\cellcolor{gray!25} &\cellcolor{gray!25} &\cellcolor{gray!25} & \cellcolor{gray!25}Accuracy, EOD,  & \cellcolor{gray!25}\\
          & \cellcolor{gray!25}~\citep{wang2023mitigating}&\cellcolor{gray!25}\multirow{-2}{*}{Partially} &\cellcolor{gray!25}\multirow{-2}{*}{I.I.D.} &\cellcolor{gray!25} \multirow{-2}{*}{In-processing}& \cellcolor{gray!25}EOP  & \cellcolor{gray!25}\multirow{-2}{*}{Active learning}\\

         \bottomrule
        \toprule
        \multirow{6}{*}{Adversarial Learning} & ARL&  \multirow{2}{*}{Completely}&\multirow{2}{*}{I.I.D.} &\multirow{2}{*}{In-processing} & \multirow{2}{*}{AUROC} & \multirow{2}{*}{Adversarial reweighting}\\
          & ~\citep{lahoti2020fairness} & &   &&   & \\
          
        \cmidrule{2-7}
        
          & \cellcolor{gray!25}Sure& \cellcolor{gray!25}&\cellcolor{gray!25} &\cellcolor{gray!25} & \cellcolor{gray!25}  &\cellcolor{gray!25}  \\
          & \cellcolor{gray!25}~\citep{chakrabarti2023sure}& \cellcolor{gray!25}\multirow{-2}{*}{Completely}&  \cellcolor{gray!25} \multirow{-2}{*}{I.I.D.}& \cellcolor{gray!25}\multirow{-2}{*}{In-processing}& \cellcolor{gray!25}\multirow{-2}{*}{Worst Group Accuracy} & \cellcolor{gray!25}\multirow{-2}{*}{Upweight samples identified with statistical significance} \\  
          
        \cmidrule{2-7}
        
          &  FairGNN & \multirow{2}{*}{Partially}&\multirow{2}{*}{Non-I.I.D.} &\multirow{2}{*}{In-processing} & Accuracy, AUROC,  & \multirow{2}{*}{Encoder-based sensitive attribute estimator} \\
          & ~\citep{dai2021say} &  &  & & EOD, SPD & \\
         \bottomrule
        \toprule
        \multirow{4}{*}{Third Parties} & \cellcolor{gray!25}FairDA &\cellcolor{gray!25} &\cellcolor{gray!25} &\cellcolor{gray!25} & \cellcolor{gray!25}Accuracy, F1-score & \cellcolor{gray!25} \\
          &  \cellcolor{gray!25}~\citep{liang2023fair}& \cellcolor{gray!25}\multirow{-2}{*}{Partially}&\cellcolor{gray!25}\multirow{-2}{*}{I.I.D.} & \cellcolor{gray!25}\multirow{-2}{*}{In-processing}& \cellcolor{gray!25} EOP, SPD  &\cellcolor{gray!25}\multirow{-2}{*}{Domain adaptation} \\
          
        \cmidrule{2-7}
         & MTL-Fair& \multirow{2}{*}{Partially}&\multirow{2}{*}{I.I.D.} &\multirow{2}{*}{In-processing} & \multirow{2}{*}{Accuracy, EOP} & \multirow{2}{*}{Multi-task learning}\\
          & ~\citep{aguirre2023transferring} & & &   &  &\\
          \bottomrule
          \toprule
        \multirow{6}{*}{Individual Fairness} & \cellcolor{gray!25}FDCPH &\cellcolor{gray!25} &\cellcolor{gray!25} &\cellcolor{gray!25} & \cellcolor{gray!25}C-index, Brier Score, AUC, &  \cellcolor{gray!25}\\
          &  \cellcolor{gray!25}~\citep{keya2021equitable}& \cellcolor{gray!25}\multirow{-2}{*}{Completely }&\cellcolor{gray!25}\multirow{-2}{*}{I.I.D.} &\cellcolor{gray!25}\multirow{-2}{*}{In-processing} &\cellcolor{gray!25}  FDCPH, FCPH  &\cellcolor{gray!25} \multirow{-2}{*}{Censorship}\\
          
        \cmidrule{2-7}
        
         & fairIndvCox& \multirow{2}{*}{Completely}&\multirow{2}{*}{I.I.D.} &\multirow{2}{*}{In-processing} & C-index, Brier score & \multirow{2}{*}{Censorship}\\
          & ~\citep{zhang2023individual} & & &   &Time-dependent, FNDCG  &\\
          
          \cmidrule{2-7}
          
         &\cellcolor{gray!25} GEIF&\cellcolor{gray!25} &\cellcolor{gray!25}  &\cellcolor{gray!25} &\cellcolor{gray!25} Accuracy, F1-score  &\cellcolor{gray!25}\\
          & \cellcolor{gray!25}~\citep{wang2024individual} & \cellcolor{gray!25}\multirow{-2}{*}{NA}&\cellcolor{gray!25}\multirow{-2}{*}{Non-I.I.D.} & \cellcolor{gray!25} \multirow{-2}{*}{In-processing} & \cellcolor{gray!25} IRGF, $\rm{IG^2F}$& \cellcolor{gray!25} \multirow{-2}{*}{Dual Fairness Objectives}\\
          
           \bottomrule
        \toprule
        \multirow{2}{*}{Fairness Under Unawareness} & FairUN&  \multirow{2}{*}{Completely}&\multirow{2}{*}{I.I.D.} &\multirow{2}{*}{Pre-processing} &Accuracy, F1-score & \multirow{2}{*}{Counterfactual fairness}\\
          & ~\citep{cornacchia2023auditing} & &   &&  SPD, EOD & \\
        \bottomrule
        \bottomrule
    \end{tabular}
    \end{adjustbox}
    \label{tab:example}
\end{table*}

\subsection{Proxy Fairness}

The proxy fairness concept originates from group fairness ideas and applies when real demographic information is unavailable or cannot be used directly. Most current research categorizes proxy fairness into two main types: one uses specific substitute demographic information~\citep{grari2021fairness}, and the other relies on unspecified or automatically inferred attributes~\citep{pelegrina2023statistical}. In the first approach, if demographic information is missing due to legal or privacy restrictions, the model relies on features that correlate with this demographic information, \textit{i.e.}, ``proxy demographic information'' inferred by a dedicated prediction model, to measure or enforce fairness. For example, variables that are strongly correlated with race or gender, or ``gender labels'' predicted by a separate classifier, can be used as proxies to approximate group fairness in the model. 

In the second approach, the idea is that the choice of predetermined demographic information is not entirely objective. These attributes are often selected beforehand and may not align with the actual dataset distribution, which can lead to outcomes that diverge from those based on the true demographic information. For example, age might not serve as meaningful demographic information in an education dataset where most students have similar ages. Some researchers~\citep{benthall2019racial} also suggest that other variables may be more appropriate for fairness analysis than standard demographic information, namely those more closely tied to the root causes of disparities. For instance, in the Gender Shades study on facial analysis algorithms, the authors avoid using race and ethnicity labels, because these are unstable (changing over time and location), and phenotypic traits can vary widely within a single category~\citep{buolamwini2018gender}. Instead, they employ skin type as a visual label. As another example, in auditing toxicity classifiers, dialect may be the most important feature~\citep{welbl2021challenges}, which highlights that the choice of categorization depends on the underlying purpose; dialect might be more relevant to system performance, whereas a legal audit could require examining potential discrimination based on demographics (\textit{i.e.}, protected attributes).

By applying group fairness notions to these proxies (\textit{e.g.}, inferred or identified features), one can approximate fairness with respect to the actual demographic information. However, if a proxy feature has a weak or incorrect correlation with the true demographic information, the model’s bias may be under- or overestimated. Both methods demand careful selection of features that best capture unobserved demographic information, with the ultimate goal of accurately measuring or mitigating bias in the model.

\subsection{Individual Fairness}


Individual fairness provides an alternative to group-based fairness metrics by emphasizing consistency at the individual level~\citep{dwork2012fairness}. Specifically, it ensures that similar individuals, defined based on input features, receive similar predictions regardless of their demographic group memberships~\citep{wang2024individual1}. Unlike group fairness definitions, individual fairness does not require predefined demographic groups or explicit demographic information, making it inherently suitable for applications where demographic information are either unavailable or intentionally omitted for privacy reasons. Formally, individual fairness can be expressed through the Lipschitz condition~\citep{dwork2012fairness}:

\begin{equation}
\label{equ:lipschitz}
D(f(x_i), f(x_j)) \leq LD'(x_i, x_j), (i \neq j)
\end{equation}

\noindent where $L$ is the Lipschitz constant, $D'(\cdot, \cdot)$ denotes the distance in input space, and $D(\cdot, \cdot)$ represents the distance between probability distributions over class labels produced by the prediction function $f(\cdot)$. Despite its intuitive appeal, applying individual fairness involves several practical and theoretical challenges. For instance, defining an appropriate similarity metric $D'(\cdot,\cdot)$ is non-trivial and context-dependent, often requiring domain knowledge or expert input to meaningfully quantify the similarity between individuals. Additionally, individual fairness constraints can indirectly lead to subgroup disparities because uniformly enforcing consistency across individuals may disproportionately influence or restrict the predictions for certain subgroups more than others~\citep{wang2024individual}. This unintended consequence suggests that only enforcing individual fairness without considering group-level outcomes could potentially exacerbate existing biases or disparities between demographic groups.

\subsection{Fairness Under Unawareness}

Fairness under unawareness refers to a fairness criterion wherein demographic information is intentionally excluded from the decision-making or modeling process. Under this definition, a model or system is considered fair if it does not explicitly utilize protected attributes when making predictions or decisions~\citep{chen2019fairness}. In contrast, fairness with incomplete demographics deals with situations where demographic information may not be deliberately excluded but is instead genuinely missing, inaccessible, or restricted by regulation~\citep{fabris2023measuring}. Thus, the primary distinction is conceptual: fairness under unawareness emphasizes intentional attribute omission, while fairness with incomplete demographics addresses inherent uncertainty or genuine unavailability of demographics.

\section{Fairness Methods Beyond Complete Demographics}
\label{sec:methods}

\begin{figure}[h]
	\centering
	\includegraphics[width=0.48\textwidth]{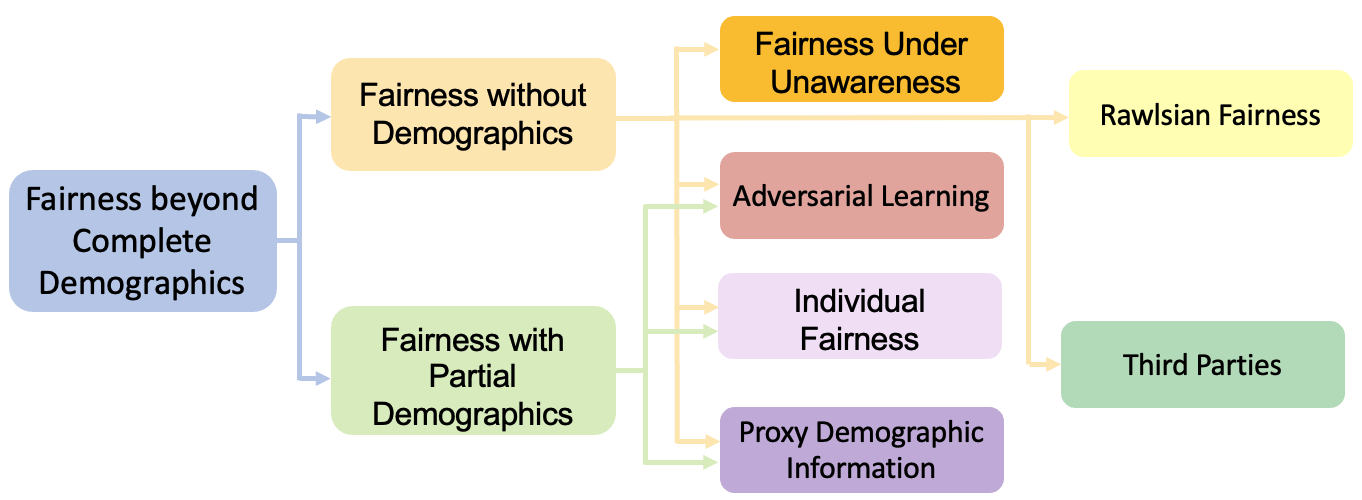}
        \vspace{+0.2cm}
	\caption{A taxonomy of the commonly used techniques to improve fairness with incomplete demographic information.}
      \vspace{+0.3cm}
	\label{fig:Taxonomy}
\end{figure}

This section reviews existing techniques for addressing fairness under varying levels of demographic missingness, providing a comprehensive overview in Table~\ref{tab:example}, where each method is analyzed across five dimensions: method, missingness, domain, class, and evaluation metrics. Specifically, the \textit{method} dimension, based on the taxonomy in Figure~\ref{fig:Taxonomy}, categorizes fairness approaches beyond complete demographics into six groups: Rawlsian fairness, adversarial learning, proxy demographic information, third-party, individual fairness and fairness under unawareness. The \textit{missingness} dimension indicates whether a method addresses partially available or completely missing demographic information, while \textit{domain} differentiates between methods designed for I.I.D. data, where samples are independent, and Non-I.I.D. data, where dependencies exist, such as in graph structures. The \textit{class} dimension specifies when fairness is enforced, distinguishing Pre-processing methods that modify or augment data before training from In-processing methods that incorporate fairness constraints during training. Finally, \textit{evaluation metrics} highlight how each method is assessed.

\vspace{-0.1cm}
\subsection{Rawlsian Fairness}

Rawlsian-based approaches follow the Max-Min Fairness principle, which minimizes the model loss for the least advantaged subgroup~\citep{hashimoto2018fairness}. Although Rawlsian methods do not require demographic information, they can leverage it when available (\textit{i.e.}, partially or completely available). For instance, groupDRO~\citep{sagawa2019distributionally} improves model performance for groups that incur a higher loss. Unlike traditional Empirical Risk Minimization (ERM), groupDRO does not optimize the average loss. However, if demographic information is missing, ERM does not explicitly address worst-off groups. To handle this, Hashimoto et al.~\citep{hashimoto2018fairness} propose distributionally robust optimization (DRO), which uses a $\chi^2$-divergence to discover and minimize the worst-case distribution repeatedly, effectively concentrating on the most disadvantaged groups. Recently, DRO-COX~\citep{hu2022distributionally} extends this idea to enhance fairness in survival analysis, where the goal is to model the time until a critical event occurs (\textit{e.g.}, death or patient recovery). Essentially, DRO acts as a reweighting mechanism that up-weights samples with higher loss. However, this approach can be overly sensitive to outliers, causing the model to focus on noisy data rather than real disadvantaged subgroups. To this end, inspired by Pareto optimality~\citep{luc2008pareto}, Martinez et al.~\citep{martinez2021blind} propose Blind Pareto Fairness (BPF), which aims to improve worst-case performance for all sufficiently large groups, ensuring that no alternative solution exists with lower group risks across all groups. This strategy provides a safeguard so that improvements to disadvantaged groups do not come at a steep cost to the best-performing groups.

A different set of methods relies on an auxiliary model (\textit{e.g.}, classifier) to identify samples from disadvantaged groups, \textit{i.e.}, use misclassified samples from the initial model as a proxy for the worst-case group, then enforces fairness by reweighting those samples. For instance, WGA~\citep{tiwari2024using} trains a reference classifier on earlier network layers, which is more effective at detecting bias-conflicting samples. In other words, if a sample is misclassified by most of the reference classifiers, it is likely to belong to the worst-case group. These samples are assigned higher weights based on how many reference classifiers misclassify them. However, this can significantly increase the computational cost due to the need to train multiple reference classifiers. It may also cause the model to pay too much attention to the worst-case group, which may lead to a decrease in the model's generalization ability on the overall dataset.

The last line of work identifies worst-off groups using the gradient magnitude or the reference model’s confidence. For example, ECS~\citep{zhao2023combating} follows a two-step procedure to address bias. First, it trains using samples with high-confidence predictions (often from the majority group). Second, it assigns sample weights according to a biased auxiliary model’s prediction probability, upweighting samples with lower confidence (often from minority groups). This encourages gradient alignment among different groups, thereby reducing bias. Conceptually, this formulation aligns with group fairness; however, in practice, focusing solely on worst-case performance does not guarantee improved inter-group fairness metrics. Additionally, a common drawback is that Max-Min fairness requires similar utility across groups, potentially leading to a noticeable drop in overall accuracy. FairKD~\citep{chai2022fairness} addresses this with a knowledge distillation (KD) approach: it first overfits a teacher model on the training set, then guides a student model using the teacher’s soft labels. By giving higher weights to ``hard'' but already correctly classified samples, the soft-label training reweights data, helping reduce disparities across groups.

Notably, all the above approaches focus on Rawlsian fairness rather than statistical parity, and experiments indicate that targeting worst-case performance does not necessarily improve parity~\cite{chai2022fairness}. At the same time, some research finds that better worst-case performance can also boost group fairness metrics such as equalized odds~\cite{lahoti2020fairness,chai2022fairness}. In addition, defining the least advantaged groups is itself challenging if the relevant criteria are not clearly specified. Nonetheless, Rawlsian fairness can be especially helpful in tackling intersectional fairness, where the model may perform poorly for subgroups formed by intersecting demographics (\textit{e.g.}, gender and nationality).

\vspace{-0.1cm}
\subsection{Adversarial Learning}

Adversarial learning has been applied to ensure fairness when demographic information is partially missing. Typically, such a framework involves a generator and a discriminator. The generator produces probabilistic predictions, while the discriminator attempts to infer demographic information from these outputs. The key idea is to set up a min-max game between the generator and the discriminator: if the discriminator fails to predict the demographic information correctly, the generator’s outputs are considered disentangled from the demographic information. In the case of partially missing demographic data, to make effective use those valuable information, an additional discriminator is introduced to estimate the missing demographic information. Thus, the entire architecture includes: i) a demographic estimator that infers missing demographic information, ii) a separate adversarial discriminator that tries to predict known or estimated demographic information, and iii) a classifier for the main label prediction task that is trained to fool the adversarial discriminator. For example, FairGNN~\citep{dai2021say} learns a fair graph neural network when only some demographic information is available. Specifically, FairGNN uses a GCN-based demographic information estimator to predict missing demographic information, then a classifier predicts node labels. This approach provides a way to apply adversarial learning, where one adversary attempts to predict known or estimated demographic information from node representations learned by the generator, while the generator learns fair node representations that make the adversary’s predictions inaccurate. Theoretical analysis shows that under mild conditions (\textit{i.e.}, the proportion of missing demographic information is not too high, so the demographic estimator can be trained accurately), this min-max game ensures that the learned representations are fair. In addition to enforcing fairness in the representations, a regularization term is added to the classifier’s predictions to further ensure fair outcomes. However, this method is not suitable if the proportion of missing demographics is too high or if the attributes are completely missing.

For cases with completely missing demographic information, researchers have developed adversarial learning approaches to achieve Rawlsian max-min fairness. For instance, ARL~\citep{lahoti2020fairness} uses ``computationally identifiable groups,'' which are regions in the input space where the model tends to make more errors. The method applies an adversary that assigns higher weights to samples that the classifier misclassifies more often, and the learner then minimizes a weighted loss. Intuitively, by focusing on these misclassified samples, the classifier is encouraged to reduce overall bias even without knowing the actual demographic information. However, this approach relies on accurate subgroup identification, which may not always be feasible. Chakrabarti~\citep{chakrabarti2023sure} argues that ARL may struggle to track all subgroups fully because of variations in unfairness patterns across training phases and datasets. To address this, the author proposes Sure~\citep{chakrabarti2023sure}, which selects a region in the latent space where the model’s error rate is statistically significant and upweights samples in that region, even correctly classified ones, in the next training iteration. The focus on the statistical significance of unfairness helps avoid upweighting misclassifications caused by random fluctuations during training. If no region shows a significant unfairness risk, training proceeds without sample reweighting. To identify subgroups, the latent representations of correctly and incorrectly misclassified samples in the current iteration are binned, and the subgroup with the largest classification error among all bins is chosen. To ensure the subgroup is sufficiently large and to handle group intersections, the bin is merged with samples from adjacent bins. However, it cannot guarantee fairness for specific demographic groups.

\vspace{-0.1cm}
\subsection{Proxy Demographic Information}

The core idea behind proxy-based methods is that proxy demographic information can be obtained when certain features correlate with the actual demographic information, or when partial demographic information is available. In scenarios where direct access to demographic information is impossible, proxy fairness notions rely on these correlated or predicted features to approximate the real demographic information. Specifically, in group fairness metrics, the demographic information can be replaced by proxy-demographic information to assess fairness. By evaluating a chosen fairness notion using these proxies, one can approximate the fairness of models with respect to the true demographic information. For example, SRCVAE~\citep{grari2021fairness} proposes a Causal Variational Autoencoder that uses a predefined causal graph to separate variables that are unaffected by the demographic information from those that depend on it. Through Bayesian inference, SRCVAE approximates the posterior and learns a latent representation enriched with demographic information. This latent variable $z$ then serves as a proxy for demographic information in an adversarial training setup, enforcing fairness criteria (\textit{e.g.}, Demographic Parity or Equalized Odds) by reducing the correlation between the prediction and $z$. However, this approach relies on a predefined causal graph to identify features correlating with the unobserved demographic information, and it is difficult to gauge the degree of correlation without domain expertise. In addition, Themis~\citep{wang2025towards} decomposes the different causal effects into multiple latent variables, allowing the separation of demographic-related information from label-related information. With this disentangled structure, demographics-related information is effectively prevented from leaking into label prediction, ensuring that the inference of demographic information remains unaffected by noisy inputs. On the other side, fairGNN-WOD~\cite{ijcai2025p63} addresses this problem through a two-stage framework to prevent optimization exploitation. By separating demographic inference from fairness-constrained learning, the model prevents manipulation of demographic assignments while preserving task-relevant demographic information. Another way is to generate proxy demographic information through clustering. For instance, KSMOTE~\citep{yan2020fair} uses clustering to assign proxy demographic labels to each group and then balances subgroup representation via oversampling, which can boost model fairness. On the other side, NOCCO~\citep{pelegrina2023statistical} proposes a preprocessing framework to identify which features are likely to yield unfair, disparate outcomes without the need to train a full predictive model. Their approach leverages the Hilbert-Schmidt independence criterion (HSIC) to measure the statistical dependence between each feature and the outcome label. If this dependence is high, it suggests that the feature provides crucial discriminative information for class prediction, likely splitting the data into subgroups with differing outcomes—thus indicating potential unfairness. Nonetheless, this method cannot guarantee that cluster labels truly reflect the demographic information of interest. In addition, FairRF~\citep{zhao2022towards}: Develops fair classifiers by exploring feature-related biases, eliminating the need for demographic information. However, this method cannot guarantee that its inferred demographic patterns accurately reflect true demographic groups.

Different from statistical fairness measures model bias, Fairwos~\citep{wang2024towards} extends counterfactual fairness to settings without observed demographic information. First, it infers proxy demographic information by learning node representations from the graph structure and non-demographic information, thereby approximating the effect of real demographic labels. Next, it seeks realistic graph counterfactuals, specific nodes, or subgraphs from the dataset that serve as plausible counterfactual scenarios while avoiding those that would be impractical in reality. By aligning the node embeddings from the original data with those from the corresponding counterfactuals, Fairwos encourages fairer predictions. Still, the inferred demographic information may not perfectly match the actual demographic information.

On the other hand, when only partial demographic information is available, this scenario is often termed a ``demographic-scarce'' regime. Here, one can leverage the known demographic labels to improve the accuracy of the proxy demographic information. For instance, FairDSR~\citep{kenfack2023fairness} proceeds in two stages: first, it trains an uncertainty-aware attribute classifier using self-ensembling with Monte Carlo dropout to predict missing demographic labels on unlabeled samples, improving both predictive accuracy and reliability of the uncertainty estimates. Similarly, DFGR~\cite{wang2025Fairness} uses limited demographic labels to train a causal encoder that produces informed demographic proxies, then applies three constraints that reduce demographic information in node representations while preserving task-relevant information. An adaptive confidence strategy weights the fairness loss by the certainty of each proxy, which limits accuracy loss while still reducing disparities across subgroups. The target classifier is then trained with fairness constraints applied only to samples whose predicted demographic labels have low uncertainty, preventing excessive performance drops from enforcing fairness on noisy or highly uncertain labels. Similarly, CGL~\citep{jung2022learning} trains an auxiliary group classifier on partially labeled data, assigning pseudo-group labels to unlabeled samples based on the classifier’s confidence; confident predictions are used as labels, while uncertain samples are assigned random labels to relax fairness constraints. Meanwhile, FDVAE~\cite{ijcai2025p64} explicitly establishes a disentanglement structure that is guided by the limited demographic labels. By doing so, it improves the recovery of missing demographic labels while preventing the model from exploiting demographic assignments to artificially improve fairness metrics. In addition, SLSD~\citep{islam2024fairness} achieves group fairness in a target domain where demographic information is unavailable by leveraging a source domain that contains these demographic labels to learn correlated latent representations for both domains. Next, it refines the group estimator on the target domain via a consistency regularization strategy that encourages robust group predictions under small input perturbations. In addition, Wang et al.~\citep{wang2023mitigating} introduce APOD, an interactive framework designed to mitigate algorithmic bias when only a small subset of training instances has demographic information. First, APOD incorporates a fairness regularizer into model training to penalize disparate outcomes between different sensitive groups. Then, through an active instance selection procedure, APOD adaptively identifies and annotates instances from the most underrepresented or misclassified subgroups, thereby exposing biases in the training data. However, this method relies on manually labeling missing demographic information and does not guarantee the accuracy of the demographic information obtained.

Beyond independent and identically distributed (i.i.d.) data, FairAC~\citep{guo2023fair} extends this concept to graph data. It first embeds nodes with observed attributes, then employs an attention mechanism to aggregate neighbor features for nodes with missing attributes. To ensure fairness, it includes a sensitive discriminator that mitigates bias both from the imputed node attributes (feature bias) and from the structural information (topological bias), producing node embeddings that are both accurate and fair. However, these methods require a sufficient number of data samples with known demographic information to train the attribute classifier effectively, and they can lose effectiveness when a large fraction of the demographics is missing.

\vspace{-0.1cm}
\subsection{Third Parties}

Instead of directly using data to identify subgroups or infer proxy demographic information, some studies propose a trusted third-party method~\citep{veale2017fairer}, which can be organized into the following three categories: First, the third party assesses the model’s fairness. In this scenario, the third party holds individual demographic information together with identifiers linking records to data or model outputs provided by the main organization. For example, MTL-Fair~\citep{aguirre2023transferring} leverages demographic labels from a related task, adapting single-task fairness objectives to a multi-task setting. The model jointly learns task A (with only target labels) and task B (with both target and demographic labels). By sharing representations from task B’s demographic data, fairness constraints are transferred to task A. Second, the third party offers high-level information about the dataset to help the main organization develop a fairer model. For instance, the third party might hold all variables and inform the main organization that ``race is correlated with zip code at 0.8''~\citep{veale2017fairer}. This information could be used when designing the model or for discussions about mitigation with the third party. FairDA~\citep{liang2023fair} demonstrates this by training a fair classifier for a target domain without demographic information using knowledge from a source domain (held by the third party). It estimates missing demographic information in the target domain by aligning latent representations between domains through adversarial objectives while using a second adversary to enforce fairness. Third, the third party acts as a data preprocessor. The third party holds all variables and preprocesses them to maintain anonymity and reduce bias. In some proposals, the main organization hands the complete dataset to the third party for fair training~\citep{jagielski2019differentially,ashurst2023fairness}. As a concrete example, the U.S. National Institute of Standards and Technology (NIST) provides data and infrastructure to audit facial recognition systems for bias~\citep{NIST_FRVT_2020}.

However, the aforementioned approaches have several limitations. While third parties with high-quality demographic data can estimate group-based metrics accurately, this requires high response rates in voluntary settings. Trust issues further complicate this, some people do not trust third parties, while others worry about data leaks or find data submission inconvenient. In addition, implementation challenges also exist. Complex mitigation strategies involving third parties become impractical when the main organization cannot verify or debug the demographic attributes. The process creates substantial costs, including legal and administrative expenses for data analysis. Security concerns present another challenge. Sharing data with additional organizations increases security risks, especially if the third party has weaker data protection measures.

\vspace{-0.1cm}
\subsection{Individual Fairness}

Methods aimed at achieving individual fairness without complete demographic information generally rely on distance metrics or ranking perspectives to ensure fairness at an individual level. For instance, FDCPH~\citep{keya2021equitable} explicitly enforces the Lipschitz constraint described by Equation~\ref{equ:lipschitz}, directly embedding individual fairness criteria into the model training process. However, specifying the appropriate Lipschitz constant remains challenging due to the inherent differences between input and output space metrics. To address this limitation, fairIndvCox~\citep{zhang2023individual} determines individual treatment differences from a ranking perspective by comparing the consistency of similarity rankings between individuals in the input and output spaces. Nevertheless, those methods might create biases against different population subgroups because the individual fairness constraint may affect subgroups differently. To this end, GEIF~\cite{wang2024individual} extends individual fairness to graph learning settings while considering the differential impacts of fairness constraints across groups. However, it still requires explicit information about group membership, which limits its applicability in scenarios with incomplete demographic information.

\vspace{-0.1cm}
\subsection{Fairness Under Unawareness}

Fairness through unawareness stipulates that sensitive attributes should not explicitly influence the decision-making process or model training. Under this notion, a decision system is considered fair if demographics are not directly included in the data or utilized explicitly in making predictions~\cite{cornacchia2023auditing}. For instance, in hiring scenarios, omitting explicit demographic details like gender or ethnicity from resumes might align with this conception of fairness. However, this approach alone does not ensure fairness, as proxy variables correlated with demographics can still indirectly influence outcomes. Features such as geographic location may inadvertently encode sensitive information due to societal biases, leading models to implicitly utilize demographics. Consequently, although fairness through unawareness helps reduce explicit bias, it does not eliminate the risk of implicit discrimination arising from correlated proxy attributes, making it generally insufficient for achieving real fairness.

\section{Benchmark Datasets}
\label{sec:datasets}

Benchmark datasets play a crucial role in ML fairness by providing standardized environments to evaluate and compare fairness methods across diverse contexts. The selection and use of these datasets significantly influence fairness analysis, as datasets embody specific demographic distributions, types of biases, and application scenarios. Generally, fairness research employs two major categories of datasets: independent and identically distributed (I.I.D.) datasets, and non-independent and identically distributed (non-I.I.D.) datasets. To facilitate clarity and comparison, Table~\ref{tab:iid_dataset_stats} summarizes widely used I.I.D. datasets across various domains, while Table~\ref{tab:niid_dataset_stats} details commonly utilized non-I.I.D. datasets represented as relational graphs.

Table~\ref{tab:iid_dataset_stats} highlights popular I.I.D. fairness datasets spanning multiple application domains, including education, criminal justice, financial services, healthcare, image recognition, social media, and other miscellaneous fields. Each dataset listed contains key statistical information such as the number of samples, features, and demographic attributes available for fairness evaluations. For instance, datasets like LSAC, COMPAS, and Adult have been extensively employed in fairness analyses involving binary classification tasks related to education outcomes, recidivism predictions, and income predictions, respectively. Similarly, image recognition datasets such as CelebA and UTKFace have facilitated studies into visual biases, leveraging attributes like gender and race. These datasets enable comprehensive fairness evaluation through defined demographics, serving as critical resources for understanding and mitigating biases in I.I.D. data.

\vspace{+0.2cm}
\begin{table}[h]
    \centering
    \caption{Summary of I.I.D. datasets used in fairness research
.}
    \vspace{+0.2cm}
    \begin{adjustbox}{width=0.43\textwidth}
    \begin{tabular}{l|llll}
        \toprule
        \textbf{Application}& \textbf{Datasets} & \textbf{\#Samples} & \textbf{\#Features} & \textbf{Demographic(s)} \\
        \midrule
       \multirow{1}{*}{Education} 
       & LSAC~\citep{wightman1998lsac}            & 27,478     & 26   & race, gender \\
        \midrule
        \multirow{2}{*}{Criminal}&COMPAS~\citep{larson2016compas}             & 7,215     & 11       & race, gender \\
        \cmidrule{2-5}
       & Violent Crime~\citep{larson2016we}        & 4,010  & 11   & race \\
         \midrule
       \multirow{5}{*}{Financial} & Default~\citep{yeh2009comparisons}           & 30,000  & 23   & gender, age \\
        \cmidrule{2-5}
        & ACSIncome~\citep{ding2021retiring}        & 1,599,299  & 10   & age, race, gender \\
        \cmidrule{2-5}
        &Adult~\citep{ding2021retiring}           & 49,531   & 14   & age, race, gender \\
        \cmidrule{2-5}
       & FLC~\citep{dispenzieri2012use}        & 7,874  & 6   & age, gender \\
       \midrule
        \multirow{5}{*}{Healthcare}& SUPPORT~\citep{knaus1995support}        & 9,105  & 14   & age, race, gender \\
        \cmidrule{2-5}
       & SEER~\citep{teng2024novel}        & 4,042  & 13   & age, race \\
       \cmidrule{2-5}
       & MIMIC-III~\citep{johnson2016mimic}         & 53,423  & 26   & race, gender, age \\
       \cmidrule{2-5}
       & ClinicalNotes~\citep{zhang2020hurtful}        & 30,598  & 26   & race, gender, age \\
         \midrule
      \multirow{6}{*}{Image } & CelebA~\citep{liu2015deep}            & 202,599  & 20   & gender \\ 
       \cmidrule{2-5}
       \multirow{5}{*}{Recognition}& UTKFace~\citep{zhang2017age}        & 20,000+  & 4   & race, gender \\
       \cmidrule{2-5}
       & Waterbirds~\citep{sagawa2019distributionally}        & 11,708  & 2   & backgrounds \\
       \cmidrule{2-5}
       & CIFAR10~\citep{krizhevsky2012imagenet}        & 60,000 & 4 & synthetic corruptions \\
       \cmidrule{2-5}
       & C-MNIST~\citep{arjovsky2019invariant}        & 70,000  & 5   & specific color \\
       \midrule
       \multirow{1}{*}{Social}& Twitter~\citep{elazar2018adversarial}         &18,470   &1,364   & political, opinion \\
      \midrule
      \multirow{7}{*}{Others} &   MEPS~\citep{statlog_(german_credit_data)_144}       & 1,000  & 23   & race \\
      \cmidrule{2-5}
       &   Toxicity~\citep{jigsaw-unintended-bias-in-toxicity-classification}       & 18,350  & -   & race \\
      \cmidrule{2-5}
      &MultiNLI~\citep{williams2017broad}        & 433,000  & -   & race, gender \\
        \cmidrule{2-5}
       & CivilComment-WILDS~\citep{koh2021wilds}       & 450,000  & -   & race, gender, religion \\
        \cmidrule{2-5}
       & OnlineReviews~\citep{hovy2015demographic}        & 5,390,000  & -   & gender, age \\
        \bottomrule
    \end{tabular}
    \end{adjustbox}
    \label{tab:iid_dataset_stats}
    \vspace{-0.2cm}
\end{table}

\vspace{+0.2cm}
\begin{table}[h]
    \centering
    \caption{Summary of non-I.I.D. datasets used in fairness research.}
    \vspace{+0.2cm}
    \begin{adjustbox}{width=0.43\textwidth}
    \begin{tabular}{l|lrrl}
        \toprule
        \textbf{Application}&\textbf{Dataset} & \textbf{\#Nodes} & \textbf{\#Edges} & \textbf{Demographic(s)} \\
        \midrule
     \multirow{4}{*}{Social}  &  Facebook~\citep{he2016ups}        & 1,034  & 26,749   &gender \\
       \cmidrule{2-5}
       & Pokec-z~\citep{takac2012data}        & 67,797  & 882,765   & gender, region \\
        \cmidrule{2-5}
       & Pokec-n~\citep{takac2012data}        & 66,569  & 729,129   & gender, region \\
        \midrule
      \multirow{2}{*}{Financial} & German~\citep{asuncion2007uci}            & 1,000     & 21,742       & gender \\
        \cmidrule{2-5}
       & Credit~\citep{yeh2009comparisons}           & 30,000  & 137,377   & age \\
       \midrule
       \multirow{2}{*}{Criminological}& Bail~\citep{jordan2015effect}           & 18,876   & 311,870   & race \\
        \cmidrule{2-5}
       & Recidivism~\citep{jordan2015effect}        &18,876   &321,308   & race \\
         \midrule
      \multirow{2}{*}{Others} & Occupation~\citep{qian2024addressing}        &6,951   &44,166   & gender \\
      \cmidrule{2-5}
       & NBA~\citep{dai2021say}            & 403     & 16,570    & country \\
        \bottomrule
    \end{tabular}
    \end{adjustbox}
    \label{tab:niid_dataset_stats}
\end{table}

Table~\ref{tab:niid_dataset_stats}, in contrast, presents a collection of non-I.I.D. fairness datasets characterized by graph-structured data. Such datasets reflect realistic scenarios involving explicit dependencies or relationships among individuals or entities, captured via nodes and edges. For instance, social network datasets (\textit{e.g.}, Pokec-z, Pokec-n, and Facebook), financial networks (\textit{e.g.}, German Credit and Credit datasets), and criminological networks (\textit{e.g.}, Bail and Recidivism datasets) exemplify cases where relational and structural biases arise naturally. Each dataset entry provides the number of nodes, edges, and associated demographic information crucial for evaluating fairness at a relational level~\citep{zhang2025fairness}. These datasets allow for in-depth investigations into fairness-aware graph learning methods, enabling researchers to explore biases arising from interactions, connectivity patterns, and relational dynamics inherent in non-I.I.D. contexts.

\vspace{-0.1cm}
\section{Challenges and Future Directions}
\label{sec:challenge}

\noindent \textbf{Formulating New Fairness Notions.} When demographics are incomplete, bias and discrimination in AI-based decision-making models become harder to detect and address. While our survey examines several fairness notions that work without complete demographics, these approaches have limitations. Some methods use proxy attributes that might not accurately represent true demographics. Others focus on worst-case performance but cannot guarantee fairness for specific demographic groups. Additionally, different fairness definitions might conflict with each other. Future research should explore fairness definitions that can adapt dynamically to varying levels of demographic completeness and provide theoretical guarantees about their robustness to specific types of missing information.

\noindent \textbf{Develop More and Tailored Datasets.} A comprehensive examination of fairness under incomplete demographic information requires diverse benchmark datasets. However, current datasets used for evaluating bias often simulate missing demographic information by manually removing these demographics, which may lead to mismatches in bias assessments. This method can introduce artificial biases and may not reflect how demographics might be genuinely missing. Creating datasets that capture realistic patterns of demographic non-disclosure, including correlations between missing data and other variables, would enable more accurate evaluation of fairness methods in these scenarios.

\noindent \textbf{Explaining How Unfairness Arises.} While several methods address bias when demographics are missing, we need a better understanding of how unfairness arises in these settings. With incomplete demographic information, unfairness can arise in new ways. For example, proxy attributes might introduce unexpected biases, or methods that estimate missing demographics might amplify existing data biases. When demographics are partially available, models might learn spurious correlations between available and missing demographic attributes. Future work should develop causal frameworks specifically designed to trace how incomplete demographic information propagates through different fairness interventions.

\noindent \textbf{Balancing Model Utility and Algorithmic Fairness.} Models that incorporate fairness considerations often experience reduced utility. In the context of incomplete demographic information, inferring missing attributes or identifying subgroups is necessary to achieve group fairness or Rawlsian fairness, combined with adding fairness-related regularization to the objective function, which can further impact model utility. Additionally, in adversarial learning frameworks, when the generator successfully fools the discriminator, some useful information may be removed from the embeddings or predictions, which can degrade performance in downstream tasks. Future research should develop theoretical bounds on the minimum performance degradation necessary when optimizing for fairness with different levels of demographic incompleteness.

\noindent \textbf{Leveraging LLMs to Address Incomplete Demographic Information.} LLMs have shown a remarkable ability to process various data types and identify hidden patterns across diverse domains~\cite{wang2024history}. In the context of fairness with incomplete demographic information, LLMs can serve as powerful tools for inferring missing demographics or measuring bias in settings where direct demographic data are unavailable due to legal, privacy, or practical constraints. By analyzing textual and contextual cues, such as social media posts or user-generated content, LLMs may reveal correlations suggestive of demographics, guiding targeted mitigation strategies even when ground truth labels are sparse. Future work should develop benchmark tasks specifically designed to evaluate how well LLMs can identify harmful biases without access to explicit demographics, and create guidelines for responsible demographic inference that respect privacy while enabling fairness interventions.

\vspace{-0.1cm}
\section{Conclusion}
\label{sec:conclusion}

There is a noticeable gap between real-world applications where demographic information is often incomplete or missing, and existing surveys on AI fairness, which largely focus on methods that assume full demographic information. This survey gives the first comprehensive analysis of fairness methods tailored for incomplete demographic information. We begin by examining fairness in scenarios with partial or missing demographic information, contrasting it with traditional ML fairness, which assumes complete demographics. Next, we introduce a novel taxonomy of fairness notions tailored to this setting and systematically review existing fairness concepts from multiple perspectives. Additionally, we categorize and describe current techniques for promoting fairness with incomplete demographics into four main groups. Finally, the survey highlights key challenges and open research questions in this emerging field.

\nocite{wang2023preventing,zhang2023individual,wang2023fg2an,wang2023mitigating,chinta2023optimization,chu2024fairness,yin2024improving,wang2023towards,chinta2024fairaied,wang2024individual1,doan2024fairness1,wang2024advancing,wang2024group,wang2024individual,yin2024accessible,wang2025fg,wang2025graph,wang2025fair,wang2025towards2,yin2025digital,chinta2025ai,pmlr-v267-wang25ek,wang2025Fairness,wang2025Redefining,zhang2019faht,zhang2024ai,zhang2022longitudinal,zhang2023censored,zhang2025fairness,zhang2022fairness,wang2024towards1,saxena2023missed,zhang2019fairness,zhang2020flexible,zhang2020online,zhang2020learning,zhang2021farf,zhang2021fair,zhang2023fairness,zhang2016using,zhang2018content,zhang2021autoencoder,zhang2018deterministic,tang2021interpretable,zhang2021disentangled,yazdani2024comprehensive,liu2021research,liu2023segdroid,cai2023exploring,guyet2022incremental,zhang2024fairness,wang2025FairnessT,zhang2025online,yinAMCR2025,Wang2025Unified,ijcai2025p64,ijcai2025p63,zhang2025datasets,palikhe2025towards,yin2025Uncertain,amon2024uncertain}

\vspace{-0.2cm}
\section*{Acknowledgements}

This work was supported in part by the National Science Foundation (NSF) under Grant No. 2404039.
\vspace{-0.2cm}


\bibliography{mybibfile}

\end{document}